\documentclass[prb,preprint,showpacs]{revtex4}
\usepackage{graphicx}
\usepackage{color}
\begin{document}

\title{
Anomalous magnetic phase diagram in low-carrier two-band systems 
and possible application to CeOs$_4$Sb$_{12}$
}

\author{
Yoshiki Imai
and Tetsuro Saso
}

\affiliation{
Department of Physics, Saitama University, Saitama 338-8570, Japan}

\date{\today}

\begin{abstract}
Magnetic properties under the external field are investigated in low-carrier two-band systems, which may explain the nontrivial phase boundary found in temperature vs. magnetic field diagram discovered in some materials, such as filled-skutterudite compound CeOs$_{4}$Sb$_{12}$. Analysis is made both for the periodic Anderson model with the small-dispersive $f$ band and the simplified two parabolic band model in the vicinity of the Fermi level. The magnetic susceptibilities are calculated by using the random phase approximation. 
It is shown that the maximum value of the magnetic susceptibility perpendicular to the external field is enhanced and yields the anomalous phase boundary. 
By applying the magnetic field, the phase boundary shifts to higher temperature region in the insulating state with a small band gap. 
On the other hand, the similar phase boundary also appears in the semi-metallic states, in which the structure of the density of states in the vicinity of the Fermi level and the finite temperature effect are essential. 
\end{abstract}

\pacs{71.10.-w, 71.27.+a}%

\maketitle
\section{Introduction}
The low-carrier concentration systems have attracted much attention because of the appearance of unusual magnetic and transport properties. 
Ce monopnictides, such as CeP and CeSb, are typical compounds, which show the very complicated phase diagram under the external field.~\cite{ross85,kohg96,iwas99}
Ce monopnictides are the semi-metals and the carrier concentrations are about 0.01 per Ce ion, so that the Fermi surfaces of electrons and holes are very small. 

Recently, the Pr-based filled skutterudite compounds have also been investigated actively, because of the discovery of fascinating superconductivities and magnetic orders.~\cite{baue02} 

The anomalous metal-insulator transition is observed in the Ce-based filled skutterudite CeOs$_{4}$Sb$_{12}$, where the charge density wave (CDW) or the spin density wave (SDW) transition occurs at 0.8 K with a gap opening of quasiparticle bands at low temperature region.~\cite{nami03} 
The phase boundary shifts to the higher temperature region with an increase of the external magnetic field,~\cite{suga05,nami03} which is contrasted to the usual phase boundaries, which shift to the lower temperature region by the field. 
CeOs$_{4}$Sb$_{12}$ may also be considered as a low-carrier system with the semi-metallic band structure. 

The resistivity of CeOs$_{4}$Sb$_{12}$ increases as $\rho \sim T^{-0.56}$ even above the transition temperature when no magnetic field is applied~\cite{suga05}, so that it seems that this material might be considered as an insulator. However, from the well-known relationship between the energy gaps and the lattice constants~\cite{suga05}, CeOs$_{4}$Sb$_{12}$ should belong to the semi-metallic group. Therefore, we consider that the basic band structure is a semi-metal, and the anomalous increase of the resistivity at low temperature may be due to disorder and/or magnetic impurity. 
Note that disorder can affect the transport properties of small carrier system drastically. This problem is to be studied separately in the future. 

Previously, we have investigated this anomalous metal-insulator transition by constructing the effective low energy model for CeOs$_{4}$Sb$_{12}$. We employed the semi-metallic band structure motivated by the LDA band structure calculation~\cite{hari03} and discussed magnetic properties. When the enhancement of the perpendicular susceptibility appears with an increase of the external field, the phase boundary between the metallic state and the antiferromagnetic insulating (AFI) state shifts to the high temperature region. Thus, the phase diagram observed in CeOs$_{4}$Sb$_{12}$ can be qualitatively reproduced. 

Although magnetic properties under the external field have been investigated for the Kondo insulator,~\cite{beac04,mila04,ohas04} we discuss not only the insulating systems but the semi-metallic systems in the present study, and investigate the generalized condition of the appearance of the anomalous phase boundary in detail. 
A part of Sec. \ref{sec:ceossb} of the present results has already been reported in the letter.~\cite{imai06} 

This paper is organized as follows. We construct the effective model with tight-binding band of CeOs$_{4}$Sb$_{12}$ and discuss magnetic properties under the external field in the next section. In Sec. \ref{freeelectron}, We introduce the two parabolic band model in order to consider the magnetic response in detail and discuss the general conditions to realize the anomalous phase transitions. Summary and discussions are given in Sec. \ref{sec:summary}. 

\section{Effective model for CeOs$_{4}$Sb$_{12}$}
\label{sec:ceossb}
In this section, we investigate magnetic properties of a filled-skutterudite compound CeOs$_{4}$Sb$_{12}$. It is because this material may have the semi-metallic band structure and becomes a typical low-carrier system with the anomalous phase boundary. 

The temperature-field ($T-B$) phase boundary of the filled skutterudite CeOs$_{4}$Sb$_{12}$ shifts to the higher temperature region with an increase of the external magnetic field.~\cite{suga05,nami03} 
Although such an anomalous phase boundary has been observed in quadrupole ordered states~\cite{taka80}, the magnetic susceptibility measurement~\cite{baue01} suggests the crystalline field ground state is $\Gamma_{7}$, which has no quadratic moment. Note that although the Sb$_{12}$ cluster has a ${\rm T_{h}}$ symmetry, $J=5/2$ ${f^{1}}$ state of the Ce atom in the Sb$_{12}$ cluster feels only the ${\rm O_{h}}$ part of the crystal field. Some experiments suggest that SDW transition is realized in this material at $T_{\rm N}=0.8$ K, below which the material changes from a metal to an insulator.~\cite{suga05} 

For the filled-skutterudite compounds, rare earth ions placed in the center of the Sb$_{12}$ clusters form body-centered cubic (BCC) lattice structure, where the $f$ orbitals on the rare-earth ions strongly hybridize the $p$ orbitals on ${\rm X_{12}}$ clusters. These $p$ orbitals form wide conduction bands. We employ the periodic Anderson model consisted of the dispersive $f$ band and the topmost of the conduction bands. The Hamiltonian is as follows: 
\begin{eqnarray}
H&=&\sum_{{\mathbf k}\sigma}\epsilon^{c}_{\mathbf k}
c^{\dag}_{{\mathbf k}\sigma}c_{{\mathbf k}\sigma}
+\sum_{{\mathbf k}\sigma}\epsilon^{f}_{\mathbf k}
f^{\dag}_{{\mathbf k}\sigma}f_{{\mathbf k}\sigma}\nonumber \\
&+&\sum_{{\mathbf k}\sigma}
\bigg(V_{\mathbf k} c^{\dag}_{{\mathbf k}\sigma}
f_{{\mathbf k}\sigma}+h.c\bigg)
+U\sum_{}n^{f}_{i\uparrow}n^{f}_{i\downarrow}\nonumber \\
&+&B_{\rm ext}\sum_{{\mathbf k}\sigma}
\sigma \bigg( n^{c}_{{\mathbf k}\sigma}
             +\beta n^{f}_{{\mathbf k}\sigma} \bigg), 
\label{ham}
\end{eqnarray}
where $c^{\dag}_{{\mathbf k}\sigma}$ ($f^{\dag}_{{\mathbf k}\sigma}$) and $c_{{\mathbf k}\sigma}$ ($f_{{\mathbf k}\sigma}$) are creation and annihilation operators of a conduction ($f$) electron with the pseudo-spin $\sigma$ (expressing $\Gamma_7$ doublet) and momentum ${\bf k}$. $\epsilon^{c}_{\mathbf k}$ ($\epsilon^{f}_{\mathbf k}$) represents the band dispersion of conduction ($f$) electrons. For the filled-skutterudite compounds, we take the BCC tight-binding bands with the nearest and the next-nearest hoppings, $\epsilon^{c}_{\mathbf k}=\alpha_{\rm c}t_{\mathbf k}$ and $\epsilon^{f}_{\mathbf k}=\alpha_{\rm f}t_{\mathbf k}+E_{f}$, where $t_{\mathbf k}=\cos(k_{x}/2)\cos(k_{y}/2)\cos(k_{z}/2)+\alpha_{2}(\cos(k_{x})+\cos(k_{y})+\cos(k_{z}))$. $V_{\mathbf k}$ is the hybridization between $f$ and conduction electrons. Although the hybridization generally has the momentum dependence, we neglect this ${\mathbf k}$ dependence ($V_{\mathbf k}=V$), for simplicity. 
 $n^{c}_{{\mathbf k},\sigma}$ ($n^{f}_{{\mathbf k},\sigma}$) represents a number operator ($n^{c}_{{\mathbf k},\sigma}=c^{\dag}_{{\mathbf k},\sigma}c_{{\mathbf k},\sigma}$, $n^{f}_{{\mathbf k},\sigma}=f^{\dag}_{{\mathbf k},\sigma}f_{{\mathbf k},\sigma}$). $U$ is the Coulomb repulsion between $f$ electrons. The last term of the Hamiltonian represents the Zeeman effect and $B_{\rm ext}$ is the applied magnetic field along $z$ direction. $\beta$ is the effective magnetic moment, which is $5/7$ (=$g_{J}\langle J_{z}\rangle$=$5/6 \times 6/7$) for Ce. 
In this section, $\alpha_{c}$ is taken to be unity, for simplicity, so that the values of other parameters and quantities will be given relative to $\alpha_{c}$. 
 
\begin{figure}[tb]
\begin{center}
\includegraphics[width=8cm]{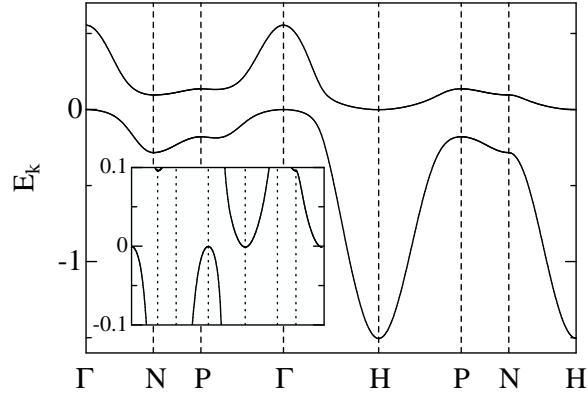}
\end{center}
\caption{Non-interacting band structure at low energy region in the absence of the external field, where $\alpha_{c}=1.0$, $\alpha_{f}=0.03$, $\alpha_{2}=0.07$, and $E_{f}=0.3$, and $V=0.15$. Fermi level lies at $\omega=0$. The inset shows the low-energy band structure. }
\label{ek1}
\end{figure}

In the non-interacting case, Hamiltonian Eq. (\ref{ham}) is easily diagonalized, whose eigenvalues are given by
\begin{eqnarray}
E^{\pm}_{\mathbf k}=\bigg(\epsilon^{c}_{\mathbf k}+\epsilon^{f}_{\mathbf k}\pm\sqrt{(\epsilon^{c}_{\mathbf k}-\epsilon^{f}_{\mathbf k})^{2}+4V^{2}}\bigg)/2. 
\label{diagek}
\end{eqnarray}
Taking account of the trend of the lattice constants of Ce-based skutterudite compounds and energy gaps, CeOs$_4$Sb$_{12}$ should belong to the semi-metallic group.~\cite{suga05} 
Figure \ref{ek1} shows the non-interacting band dispersion in the absence of external magnetic field at low-energy regime. Although we employ the simple model, the obtained band structure captures the essential features in comparison with the result of the LDA calculation~\cite{hari03} at low energy region, where the lower band at $\Gamma$ point and the upper band at H point slightly intersect the Fermi level. Our obtained band dispersions also show the semi-metallic character. 

In general, the Coulomb repulsion between $f$ electrons is screened in the rare-earth metals. Furthermore, the effects of the Hubbard-type interaction may be suppressed by the external field. 
Therefore, in order to investigate the Hamiltonian Eq. (\ref{ham}), we employ the random phase approximation (RPA), which is one of the mean-field approximations and is justified in the weak coupling regime. Then, magnetic susceptibilities for $x$ and $z$ directions are given by 
\begin{eqnarray}
&&\chi^{\pm} ({\mathbf q})=\chi^{y} ({\mathbf q})=\frac{\beta^{2}}{2}
\frac{\chi^{0}_{\uparrow \downarrow}({\mathbf q})}
{1-U\chi^{0}_{\uparrow \downarrow}({\mathbf q})},
\label{eqn:chix}
\\
&&\chi^{z} ({\mathbf q})=\frac{\beta^{2}}{4}
\frac{
     \chi^{0}_{\uparrow \uparrow}({\mathbf q})
    +\chi^{0}_{\downarrow \downarrow}({\mathbf q})
  +2U\chi^{0}_{\uparrow \uparrow}({\mathbf q})
     \chi^{0}_{\downarrow \downarrow}({\mathbf q})
     }
    {1-U^{2}\chi^{0}_{\uparrow \uparrow}({\mathbf q})
            \chi^{0}_{\downarrow \downarrow}({\mathbf q})},
\label{eqn:chiz}
\end{eqnarray}
where $\chi^{0}_{\sigma \sigma'}$ is the band susceptibility and is defined as 
\begin{eqnarray}
&&\chi^{0}_{\sigma \sigma'} ({\mathbf q})=
  \frac{-T}{N}
  \sum_{{\mathbf k},\omega_{n}}
    g^{\rm f}_{\sigma}({\mathbf k}+{\mathbf q},{\rm i}\omega_{n})
    g^{\rm f}_{\sigma'}({\mathbf k},{\rm i}\omega_{n}),
\label{bandchi0}
\\
&&g^{\rm f}_{\sigma}({\mathbf k},{\rm i} \omega_{n})
   =\frac{1}
     { {\rm i}\omega_{n} +\mu -\epsilon^{f}_{{\mathbf k}\sigma}
      -{\displaystyle \frac{V^{2}}{{\rm i}\omega_{n} +\mu -\epsilon^{c}_{{\mathbf k}\sigma}}} }, 
\end{eqnarray}
where $\epsilon^{c}_{{\mathbf k}\sigma}=\epsilon^{c}_{\mathbf k}+\sigma B_{\rm ext}$ and $\epsilon^{f}_{{\mathbf k}\sigma}=\epsilon^{f}_{\mathbf k}+\sigma \beta B_{\rm ext}$. 

Note that the band susceptibility $\chi({\mathbf q})$ consists of $4\times 4$ matrix for each momentum ${\mathbf q}$ in this treatment in an exact formulation. 
Since $f$ electrons are strongly localized in comparison with conduction electrons, the $f$-$f$ component of the band susceptibility is dominant for the effect of the external field. Furthermore, since only the $f$ electrons have the strong Coulomb repulsion, the magnetic instability mainly results from $f$ electrons.  
Hereafter, we consider only the $f$-diagonal component, for simplicity. 

\begin{figure}[tb]
\begin{center}
\includegraphics[width=7cm]{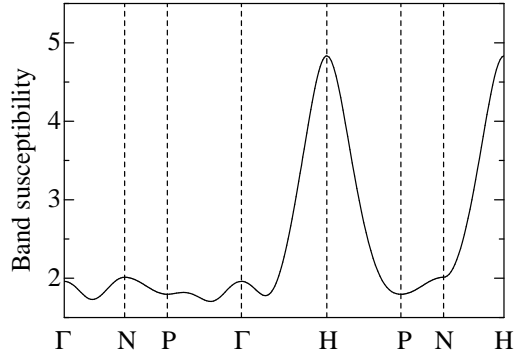}
\end{center}
\caption{Static band susceptibility for various momenta in the absence of the external field at $T=0.004$. 
}
\label{fig:chi0}
\end{figure}

Figure \ref{fig:chi0} shows the momentum dependence of the band susceptibility.  
The maximum peak appears at ${\mathbf q}=\frac{2\pi}{a}(1,0,0)$. 
Let us discuss the instability for the ordered state. In the absence of the external field, the band susceptibility $\chi^{0}_{\sigma \sigma'}$ is independent of the spin index. Considering Eqs. (\ref{eqn:chix}) and (\ref{eqn:chiz}) within RPA, the maximum peak at H point indicates that the ordering vector is ${\bf Q}=\frac{2\pi}{a}(1,0,0)$, at which $\chi^{\pm} ({\mathbf Q})$ and $\chi^{z} ({\mathbf Q})$ are strongly enhanced with an increase of the Coulomb interaction. This behavior can be explained by the nesting property. Since the Fermi surfaces become small spheres at the top of the lower band ($\Gamma$ point) and at the bottom of the upper band (H point), the particle-hole excitations can easily occur between these momenta, which can be easily understood in Fig. \ref{ek1}. The recent neutron scattering experiment for CeOs$_{4}$Sb$_{12}$ shows that the ordering vector in the absence of the external field corresponds to $\frac{2\pi}{a}(1,0,0)$.~\cite{iwas05}
Therefore, although the LDA results have a conduction band with a concave dispersion near $\Gamma$ point in addition to the band we assumed, it may not affect magnetic properties of CeOs$_{4}$Sb$_{12}$. 

\begin{figure}[tb]
\begin{center}
\includegraphics[width=7cm]{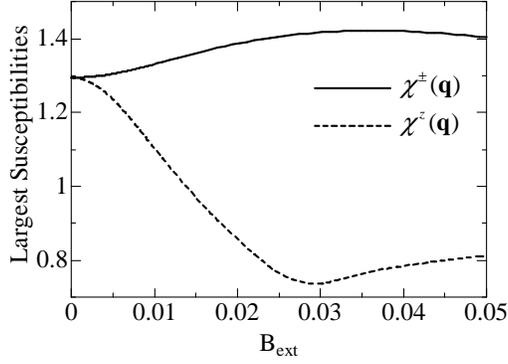}
\end{center}
\caption{Largest magnetic susceptibilities as a function of the external field at $T=0.003$ where $U=0$. The solid (dashed) line represents perpendicular (parallel) component to the external field. }
\label{fig:chih}
\end{figure}

\begin{figure}[tb]
\begin{center}
\includegraphics[width=7cm]{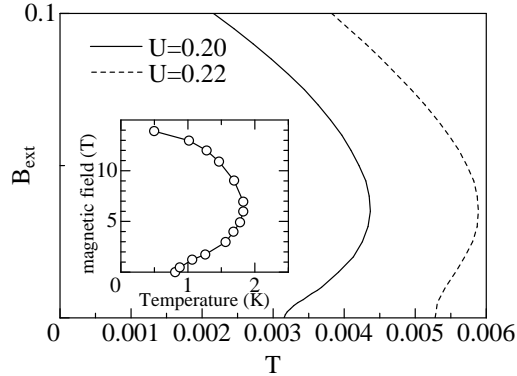}
\end{center}
\caption{Temperature-external field phase diagram at $U=0.2$ (solid) and $U=0.22$ (dashed). The inset shows the experimental result.~\cite{suga05}}
\label{fig:pd}
\end{figure}

Next, we investigate the magnetic response under the external field. Figure \ref{fig:chih} shows the largest susceptibilities for the parallel component ($\chi^{z}$) and perpendicular component ($\chi^{\pm}$) to the external field. 
For the perpendicular component ($\chi^{\pm}$), the maximum occurs at ${\mathbf Q}=\frac{2\pi}{a}(1,0,0)$, which does not change as far as $B_{\rm ext} < 0.2$. On the other hand, the peak position of the parallel component gradually changes with an increase of the external field. 
The largest value of $\chi^{\pm}$ is enhanced within small $B_{\rm ext}$ range because the external field may change the number of the states close to the Fermi level, which is discussed in Sec. \ref{freeelectron} in detail.  
This result indicates that the instability for the magnetic order is enhanced by the external field. 
The transition temperature $T_{\rm N}$ of the metal-AFI transition is determined by $1-U\chi^{0}_{\uparrow \downarrow}(T_{\rm N})= 0$ within RPA. 
Because of the enhancement of $\chi^{\pm}$, the phase boundary of the metal-AFI transition shifts to the higher temperature region with an increase of $B_{\rm ext}$, which is shown in Fig. \ref{fig:pd}. 

Note that except for $B_{\rm ext}=0$, the direction of the magnetic moments are perpendicular to the external field in the ordered state in the present study. 
In the very weak field region, the shift of the phase boundary is independent of the external field. Namely, the slope of the phase boundary as a function of temperature becomes infinite, which results from the symmetry of the Hamiltonian with respect to the reversal of the direction of the field. 
However, since magnetic anisotropies exist in the realistic materials due to the coupling to the lattice, the magnetic moment may not suddenly be directed to $x$ or $y$ axis, but only be slightly canted from $z$ direction. 

Further increasing the external field, the phase boundary is closed. The obtained anomalous $T-B$ phase diagram can quantitatively reproduce the experimental result of CeOs$_4$Sb$_{12}$, which indicates that the effective model in this section may describe qualitative properties of this material. 

\section{General Conditions for Anomalous Phase Diagram}
\label{freeelectron}
In the previous section, we have obtained anomalous $T-B$ phase diagram by constructing the effective model of CeOs$_4$Sb$_{12}$, where the enhancement of the perpendicular susceptibility $\chi^{\pm}$ with an increase of the external field was essential. 
In this section, we investigate in detail the general condition to realize the anomalous phase diagram in the two-band system with low carriers, such as the semi-metallic system shown in Fig. \ref{ek1}. 
For this purpose, we introduce a simplified model Hamiltonian. 
Since the external field dependence of the band susceptibility is directly related to the AFI-metal phase boundary within RPA, which was discussed in Sec. \ref{sec:ceossb}, hereafter we focus on the band susceptibility. 

\begin{figure}[tb]
\begin{center}
\includegraphics[width=8cm]{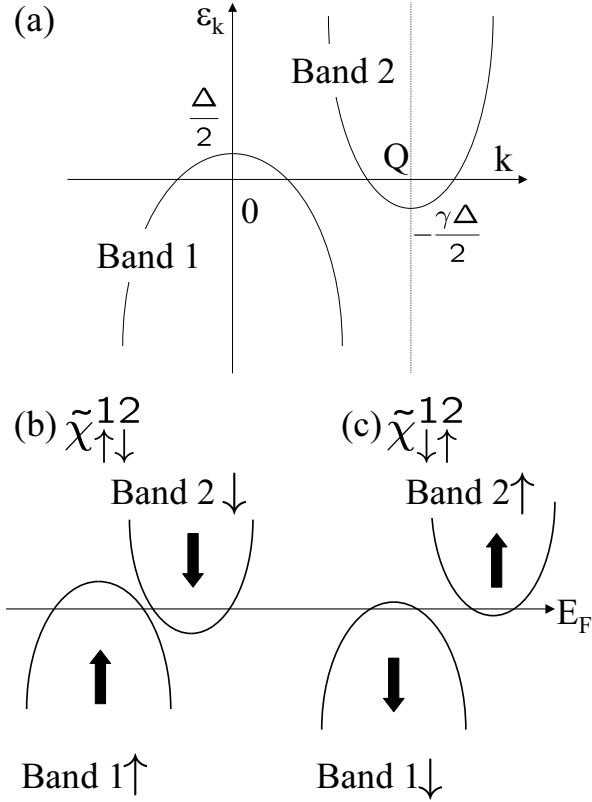}
\end{center}
\caption{(a) Schematic band dispersions in the absence of external field ($\Delta >0$) for one dimension. (b) and (c) stand for the shifts by the external field in the band off-diagonal components $\tilde{\chi}^{12}_{\uparrow \downarrow}$ and $\tilde{\chi}^{12}_{\downarrow \uparrow}$ of Eq. (\ref{eqn:chiud}) at $\Delta >0$. Bold arrows stand for the shift directions due to the external field. 
}
\label{ek2}
\end{figure}

The energy dispersions of the semi-metallic system close to the Fermi level are the almost parabolic structure with respect to momenta ${\mathbf k}$ with low carriers (Fig. \ref{ek1} inset). 
Therefore, we approximate the effective band dispersion close to the Fermi level by the two parabolic bands. Then the Hamiltonian for the band dispersions is given as 
\begin{eqnarray}
\tilde{H}&=&\sum_{{\mathbf k}\alpha \sigma}
\epsilon^{\alpha}_{{\mathbf k} \sigma} 
a^{\dag}_{{\mathbf k}\alpha\sigma}a_{{\mathbf k}\alpha\sigma}, \\
\epsilon^{\alpha}_{{\mathbf k} \sigma}&=&
\left \{
\begin{array}{c}
-\{{\mathbf k}^{2}-\frac{\Delta}{2}\}
+\sigma B_{\rm ext} \,\,\,\, (\alpha =1) \\
\gamma \{({\mathbf k}-{\mathbf Q})^{2}-\frac{\Delta}{2} \}
+\sigma B_{\rm ext} \,\,\,\, (\alpha =2), 
\end{array}
\right. 
\label{eqn:ek2free}
\end{eqnarray}
where $a^{\dag}_{{\mathbf k}\alpha \sigma}$ ($a_{{\mathbf k}\alpha \sigma}$) is creation (annihilation) operator of an electron with the pseudo-spin $\sigma$, momentum ${\bf k}$, and band index $\alpha (=1,2)$. We assume $2m=1$ for simplicity, and $\gamma$ stands for the curvature of the band 2 relative to that of the band 1 and the bottom of the band 2 is located at ${\mathbf k}={\mathbf Q}$. 
$\Delta$ stands for the overlap ($\Delta >0$) or gap ($\Delta <0$) between two bands. 
Figure \ref{ek2}(a) shows the schematic band dispersions of Eq. (\ref{eqn:ek2free}) in the absence of external field. 

Then, the perpendicular and parallel components of the band susceptibility are defined as 
\begin{eqnarray}
\tilde{\chi}^{\pm}_{0}({\mathbf q})&=&
 \sum_{\sigma}\tilde{\chi}^{12}_{\sigma \bar{\sigma}}({\mathbf q})
+\sum_{\alpha}\tilde{\chi}^{\alpha \alpha}_{\uparrow \downarrow}({\mathbf q}),
\label{eqn:chipm0}
\\
\tilde{\chi}^{z}_{0}({\mathbf q})&=&
 \sum_{\alpha \beta \sigma}\tilde{\chi}^{\alpha \beta}_{\sigma \sigma}({\mathbf q}),
\label{eqn:chiz0}
\\
\tilde{\chi}^{\alpha \beta}_{\sigma \sigma'}({\mathbf q})
&=&\frac{-1}{N}\sum_{{\mathbf k}}
\frac{f(\epsilon^{\alpha}_{{\mathbf k}+{\mathbf q} \sigma})
     -f(\epsilon^{\beta}_{{\mathbf k} \sigma'})}
     {\epsilon^{\alpha}_{{\mathbf k}+{\mathbf q} \sigma}
     -\epsilon^{\beta}_{{\mathbf k} \sigma'}}, 
\label{eqn:chiud}
\\
\tilde{\chi}^{\alpha \alpha}_{\sigma \sigma'}({\mathbf q})
&=&\frac{-1}{N}\sum_{{\mathbf k}}
\frac{f(\epsilon^{\alpha}_{{\mathbf k}+{\mathbf q} \uparrow})
     -f(\epsilon^{\alpha}_{{\mathbf k} \downarrow})}
     {\epsilon^{\alpha}_{{\mathbf k}+{\mathbf q} \uparrow}
     -\epsilon^{\alpha}_{{\mathbf k} \downarrow}}, 
\label{eqn:chidiag}
\end{eqnarray}
where $f(\epsilon)$ stands for the Fermi distribution function. 
The schematic behavior for the external field is shown in Fig. \ref{ek2}(b) and (c). 
Note that the amplitude of the band diagonal susceptibility (Eq. (\ref{eqn:chidiag})) is small in comparison with that of the band off-diagonal component under the weak external field. 

\subsection{Absolute Zero Temperature}
At absolute zero temperature ($T=0$), we can analytically calculate Eqs. (\ref{eqn:chipm0})-(\ref{eqn:chidiag}). The semi-metallic state ($\Delta > 0$) has small hole (band 1) and electron (band 2) Fermi surfaces. 
In the absence of the external field, the volume of the Fermi surface of the band 1 is equal to that of the band 2. The radius of each Fermi surface is $k^{1}_{{\rm F}\sigma}=k^{2}_{{\rm F}\sigma}=\sqrt{\frac{\Delta}{2}}$ $(\Delta >0)$, where $k^{\alpha}_{{\rm F}\sigma}$ is Fermi wave vector of band $\alpha$ with the pseudo spin $\sigma$. Therefore, the perfect nesting of the Fermi surfaces is realized, so that the band susceptibility always diverges at ${\mathbf q}={\mathbf Q}$ regardless of any spatial dimensions, where ${\mathbf Q}$ becomes the ordering vector. 
If the orbital degeneracy remains, the perfect nesting condition may vanish even in the absence of the external field. The effect of the orbital degeneracy will be discussed in later section. 

The radius of each Fermi sphere is gradually modified as the external field is applied. 
When both bands intersect the Fermi level which satisfies the following relations $\Delta/2 +\sigma B_{\rm ext} > 0$ and $\Delta/2-\sigma B_{\rm ext}/\gamma > 0$, both Fermi wave vectors at small $B_{\rm ext}$ are given by 
\begin{eqnarray}
\left \{
\begin{array}{c}
k^{1}_{{\rm F}\sigma}=\sqrt{\frac{\Delta}{2}+\sigma B_{\rm ext}} \\
k^{2}_{{\rm F}\sigma}=\sqrt{\frac{\Delta}{2}-\frac{\sigma B_{\rm ext}}{\gamma}}. 
\end{array}
\right.
\end{eqnarray}
If a larger external field is applied, band 1 with pseudospin $\downarrow$ and band 2 with pseudospin $\uparrow$ can not intersect the Fermi level, where the Fermi surfaces of band 1 with pseudospin $\downarrow$ and of band 2 with pseudospin $\uparrow$ are lost. 

If the curvature of the band 2 is equal to that of the band 1 ($\gamma=1$), the volume of Fermi sphere of the band 1 with spin $\sigma$ is always equal to that of the band 2 with spin $\bar{\sigma}$ even in the presence of the external field, where the perfect nesting still remains between different bands. 
Therefore, when the infinitesimal Coulomb interaction is introduced, the magnetic susceptibility always diverges regardless of the presence of the external field within RPA. 
Since the perfect nesting vanishes for $\gamma \ne 1$ and $B_{\rm ext} \ne 0$, the band susceptibility remains finite, so that magnetic instabilities are suppressed in comparison with the case of $\gamma =1$. 

On the other hand, if $\Delta <0$ where the band insulator with gap $|\Delta|$ is realized, the amplitude of the band susceptibility becomes small. When the gap vanishes by applying the external field, the band susceptibility $\tilde{\chi}^{\pm}_{0}({\mathbf Q})$ diverges ($\gamma =1$) and becomes enhanced to a large value ($\gamma \ne 1$). 

However, our interest is the origin of the anomalous behavior of temperature-field phase diagram. In next subsection, we discuss the magnetic response to the external field at finite temperature. 

\subsection{Low Temperature Susceptibility}
Magnetic order cannot occur at finite temperature except for the three-dimensional case. 
However, in order to clarify the origin of the anomalous phase boundary, the magnetic susceptibilities for various dimensions are investigated in detail within the mean-field level in the very low-temperature region ($T=0.01$) in this subsection. 

Since, unfortunately, it is not easy to carry out the analytical calculations of the band susceptibility at finite temperature, we numerically evaluate Eqs. (\ref{eqn:chipm0})-(\ref{eqn:chidiag}). 
Note that since we employ the parabolic band in energy dispersions, the bandwidth becomes infinite. In order to maintain the finite bandwidth, we restrict the wave-vectors within $|k_{x}|+|k_{y}|<\pi$, $|k_{y}|+|k_{z}|<\pi$, and $|k_{z}|+|k_{x}|<\pi$ in numerical calculations. However, this restriction is irrelevant for magnetic properties at low temperature and weak external field. 

The density of states (DOS) reproduces $\rho(\omega)\sim |\omega|^{d/2-1}$ at $\omega \sim 0$ for $d$ dimensions at $\Delta=B_{\rm ext}=0$, which is the well-known result. Because of the restriction of the wave-vectors, DOS rapidly decreases at sufficiently large $|\omega|$ ($>5$) region even in two and three dimensions. 

\begin{figure}[tb]
\begin{center}
\includegraphics[width=8cm]{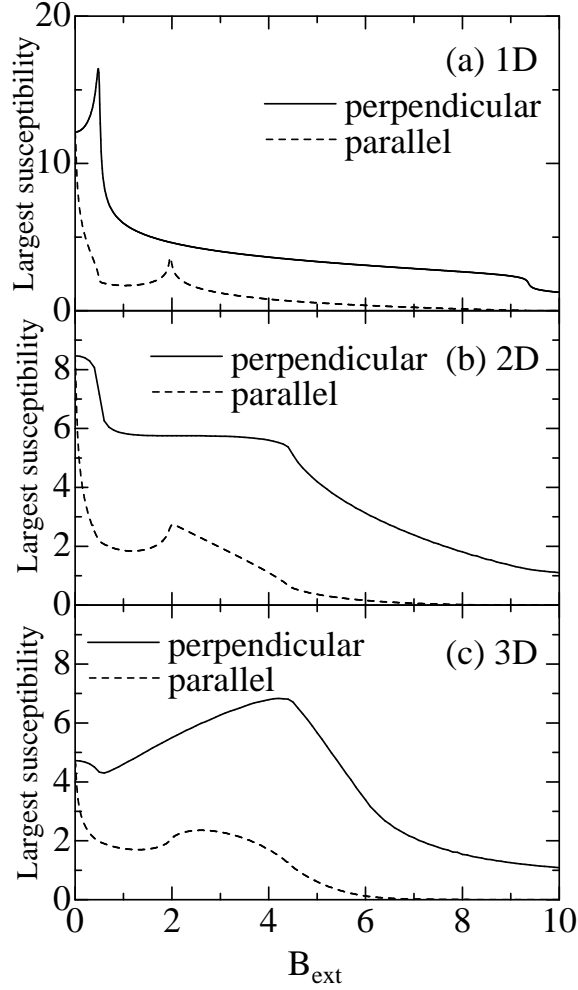}
\end{center}
\caption{Largest values of perpendicular and parallel components of band susceptibilities as a function of external field for (a) one dimension, (b) two dimensions, and (c) three dimensions at $T=0.01$, $\Delta=1$, and $\gamma=1$. The solid (dotted) lines stand for perpendicular (parallel) components. The wave vectors in all perpendicular components are ${\mathbf q}={\mathbf Q}$. }
\label{fig:chiszd}
\end{figure}

First, let us compare the amplitude of the perpendicular susceptibility $\tilde{\chi}^{\pm}_{0}({\mathbf q})$ with that of the parallel susceptibility $\tilde{\chi}^{z}_{0}({\mathbf q})$ under the magnetic field for various spatial dimensions at $\gamma=1$. 
Figure \ref{fig:chiszd} shows the largest values of the band susceptibilities at $T=0.01$ in the semi-metallic case for various dimensions. 
The perpendicular components of the band susceptibilities become always larger than those of the parallel components in whole range of the external field and all dimensions.
The largest values of the perpendicular components for all dimensions appear at ${\mathbf q}={\mathbf Q}$, which corresponds to the nesting vector, while the ordering wave vectors of the parallel components are gradually shifted as the external field increases. 
For the insulating case ($\Delta <0$), the perpendicular susceptibilities are always larger than the parallel susceptibilities. 

Since these behaviors are similar to the results of Sec. \ref{sec:ceossb}, qualitative properties of the anomalous phase-diagram obtained in Sec. \ref{sec:ceossb} may be captured even in this simplified model. 
The obtained result indicates that the magnetic instability perpendicular to the external field occurs in this system. 
Therefore, we hereafter focus on the behavior of the perpendicular susceptibility $\tilde{\chi}^{\pm}_{0}({\mathbf q})$. 

Although the perpendicular susceptibility $\tilde{\chi}^{\pm}_{0}({\mathbf q})$ consists of the band diagonal and the off-diagonal parts, the amplitude of the latter is greater than that of the former. 
Thus, magnetic properties are almost determined by the off-diagonal part, which consists of two components, $\tilde{\chi}^{12}_{\uparrow \downarrow}$ and $\tilde{\chi}^{12}_{\downarrow \uparrow}$ terms. 
The schematic behavior of the band off-diagonal susceptibilities is shown in Fig. \ref{ek2}(b) and (c), where there are contributions from the band dispersions shifting away from the Fermi level and approaching the Fermi level with an increase of the external field. 
Although the perfect nesting still exists at $\gamma=1$ even under the external field, the amplitude of the band susceptibility becomes finite due to the finite temperature effect even at ${\bf q}={\bf Q}$. 

\begin{figure}[tb]
\begin{center}
\includegraphics[width=8cm]{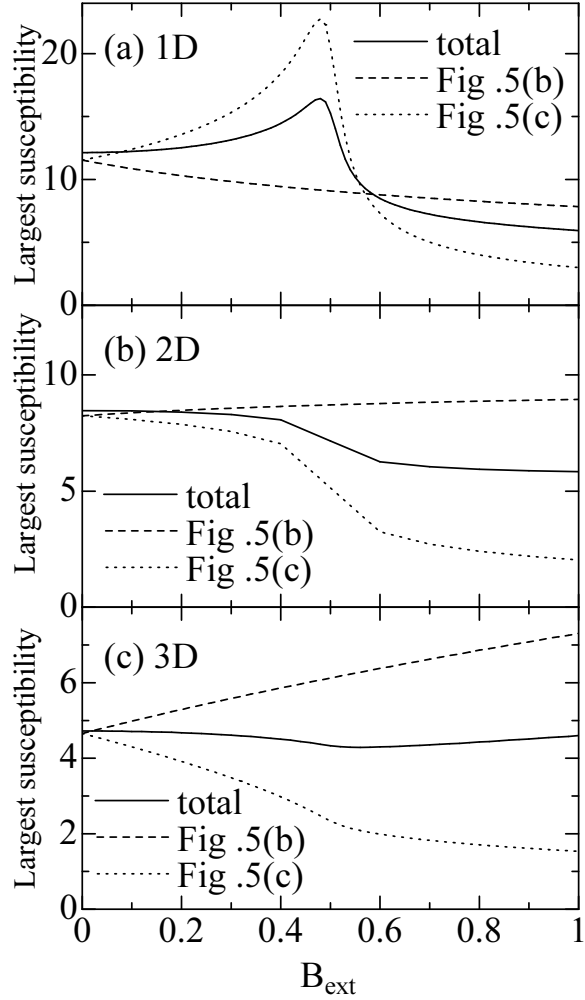}
\end{center}
\caption{
Largest values of perpendicular susceptibility $\tilde{\chi}^{\pm}_{0}$ and band off-diagonal components as a function of external field at $\Delta=1$, $T=0.01$, and $\gamma=1$ at low field region. The solid, dashed, and dotted lines stand for $\tilde{\chi}^{\pm}_{0}$ (total), $\tilde{\chi}^{12}_{\uparrow \downarrow}$ (Fig. \ref{ek2}(b)), and $\tilde{\chi}^{12}_{\downarrow \uparrow}$ (Fig. \ref{ek2}(c)), respectively. All wave vectors are ${\mathbf q}={\mathbf Q}$. 
}
\label{fig:chipart}
\end{figure}

In order to understand properties of the perpendicular susceptibilities, we show two off-diagonal components of the perpendicular susceptibilities in Fig. \ref{fig:chipart} for the semi-metallic case ($\Delta=1$). 
For one-dimensional system, $\tilde{\chi}^{12}_{\uparrow \downarrow}$ (Fig. \ref{ek2}(b)) decreases while $\tilde{\chi}^{12}_{\downarrow \uparrow}$ (Fig. \ref{ek2}(c)) increases as the small external field is applied. 
For two-dimensional system, $\tilde{\chi}^{12}_{\uparrow \downarrow}$ is almost independent of the external field, while $\tilde{\chi}^{12}_{\downarrow \uparrow}$ (Fig. \ref{ek2}(c)) is strongly suppressed at $B_{\rm ext} >\Delta/2$. 
On the other hand, $\tilde{\chi}^{12}_{\uparrow \downarrow}$ increases while $\tilde{\chi}^{12}_{\downarrow \uparrow}$ decreases in three dimensions as the external field is applied.
In addition to these behaviors, the conspicuous changes appear at $B_{\rm ext}\sim\Delta/2$ for all dimensional cases. 
This result indicates that the number of the states close to the Fermi level is directly related to the amplitude of susceptibility. 
With an increase of the external field, the number of the states close to the Fermi level for the cases Fig. \ref{ek2}(b) and (c) is modified. 
Thus, the weight of particle-hole excitations at low temperature also drastically changed by the external field, which affects the amplitude of $\tilde{\chi}^{12}_{\uparrow \downarrow}$ and that of $\tilde{\chi}^{12}_{\downarrow \uparrow}$. 

For one-dimensional system, DOS $\rho^{\alpha} (\omega)$ with band index $\alpha$ has a maximum peak on each band edge ($\omega =\pm \Delta/2$). 
Therefore, when the overlap between two bands for $\tilde{\chi}^{12}_{\downarrow \uparrow}$ (Fig. \ref{ek2}(c)) disappears for by applying $B_{\rm ext} \sim \Delta/2$, the amplitude of $\tilde{\chi}^{12}_{\downarrow \uparrow}$ becomes maximum. 
For $\tilde{\chi}^{12}_{\uparrow \downarrow}$ (Fig. \ref{ek2}(b)), the change of the number of the states becomes small in comparison with that of Fig. \ref{ek2}(c). 
The total susceptibility $\tilde{\chi}^{\pm}_{0}$ is mainly determined by the sum of $\tilde{\chi}^{12}_{\uparrow \downarrow}$ and $\tilde{\chi}^{12}_{\downarrow \uparrow}$, so that a enhancement of $\tilde{\chi}^{\pm}_{0}$ for the external field appears. 

For larger dimensions, the DOS is related to the structure of the perpendicular susceptibility $\tilde{\chi}^{\pm}_{0}$. 
In the two-dimensional system, since DOS is independent of frequency $\omega$, $\tilde{\chi}^{12}_{\uparrow \downarrow}$ almost becomes constant. 
However, the overlap between two bands for $\tilde{\chi}^{12}_{\downarrow \uparrow}$ disappears at $B_{\rm ext}>\Delta/2$, so that the amplitude is reduced at $B_{\rm ext}>\Delta/2$. Therefore the total susceptibility $\tilde{\chi}^{\pm}_{0}$ is also reduced at $B_{\rm ext}>\Delta/2$.
In the three-dimensional case, DOS for each band is $\rho^{1}_{\sigma}(\omega)=|\omega -\Delta/2-\sigma B_{\rm ext}|^{\frac{1}{2}}$ and $\rho^{2}_{\sigma}(\omega)=|\omega +\Delta/2-\sigma B_{\rm ext}|^{\frac{1}{2}}$. 
The ratio of change of the number of the states 
to the external field is approximately represented as ${\rm d}\rho(\omega)/{\rm d}B_{\rm ext}|_{\omega =0}\sim |\Delta/2+B_{\rm ext}|^{-\frac{1}{2}}$ for Fig. \ref{ek2}(b) and $|\Delta/2- B_{\rm ext}|^{-\frac{1}{2}}$ for Fig. \ref{ek2}(c) at sufficiently low temperature, 
so that the decrease of the number of the states for $\tilde{\chi}^{12}_{\downarrow \uparrow}$ becomes slightly larger than 
the increase of the number of the states for $\tilde{\chi}^{12}_{\uparrow \downarrow}$ in the small external field region. 
Thus, the total perpendicular susceptibility $\tilde{\chi}^{\pm}_{0}$ is reduced at $B_{\rm ext}<\Delta/2$ while it is enhanced with further increase of $B_{\rm ext}$ $(>\Delta/2$). 

Although the maximum of the band susceptibilities appears at the nesting vector at finite temperature, the amplitude strongly depends on the structure of DOS close to Fermi level. 
In the present study, our main objective is to explore the possibility of the anomalous phase transition for three-dimensional realistic systems. 
Hereafter, we discuss magnetic properties more in detail in three dimensions. 

\begin{figure}[tb]
\begin{center}
\includegraphics[width=8cm]{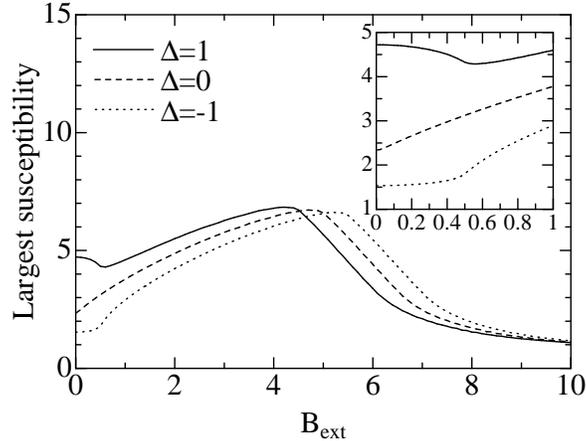}
\end{center}
\caption{Largest values of perpendicular components of band susceptibilities $\tilde{\chi}^{\pm}_{0}$ as a function of external field at $T=0.01$ and $\gamma=1$ for three-dimensional system. The solid, dashed, and dotted lines stand for $\Delta=1$ (semi-metal), $\Delta=0$, and $\Delta=-1$ (insulator), respectively. All wave vectors are ${\mathbf q}={\mathbf Q}$. The inset shows the low-field part.}
\label{fig:chixd}
\end{figure}
In Fig. \ref{fig:chixd}, we show the perpendicular components of the band susceptibilities for various $\Delta$ in the three-dimensional system. 
The cusps appear around at $B_{\rm ext} \sim 5$, which arise from the band edge due to the introductions of the cutoff wave vectors. 
If the cutoff wave vectors are absent, the amplitude gradually increases for the three-dimensional system in the present model. 

However, our aim is to clarify magnetic properties in the low field region. 
The difference in each perpendicular susceptibility particularly appears at $B_{\rm ext} < |\Delta|/2$. 
For the $\Delta<0$ case, which corresponds to a insulating state with a gap $|\Delta|$, the enhancements of the susceptibilities for the external field can be realized at $B_{\rm ext} > |\Delta|/2$. When $B_{\rm ext}=0$, there is no states close to the Fermi level at low temperature due to the existence of the insulating gap. When the gap in DOS vanishes due to the introduction of the external field ($B_{\rm ext}\sim |\Delta|/2$), $\tilde{\chi}^{12}_{\uparrow \downarrow}$ is strongly enhanced. 
Since there is no states close to the Fermi level for $\tilde{\chi}^{12}_{\downarrow \uparrow}$ in the whole field region, the amplitude becomes almost constant at low temperature. 
Therefore, as far as the insulating state ($\Delta < 0$) with a sufficiently small gap, the enhancement of the magnetic susceptibility $\tilde{\chi}^{\pm}_{0}$ can be realized in the low field region. 
On the other hand, for $\Delta>0$, each band intersects the Fermi level and the semi-metallic state is realized, where the overlap amplitude between two bands is $\Delta$ at $B_{\rm ext}=0$. 
As the explanation for $\Delta > 0$ is given in Fig. \ref{fig:chipart}, the perpendicular susceptibility $\tilde{\chi}^{\pm}_{0}$  for the external field is suppressed within $B_{\rm ext} < \Delta/2$. 

Note here that the suppression of the perpendicular susceptibility in the three-dimensional system is weakened in comparison with that of the two-dimensional system in the semi-metallic states. 
It is because that the change of the number of the states close to the Fermi level becomes gradual.

\subsection{Susceptibilities at finite Temperatures and Phase Diagram}

\begin{figure}[tb]
\begin{center}
\includegraphics[width=8cm]{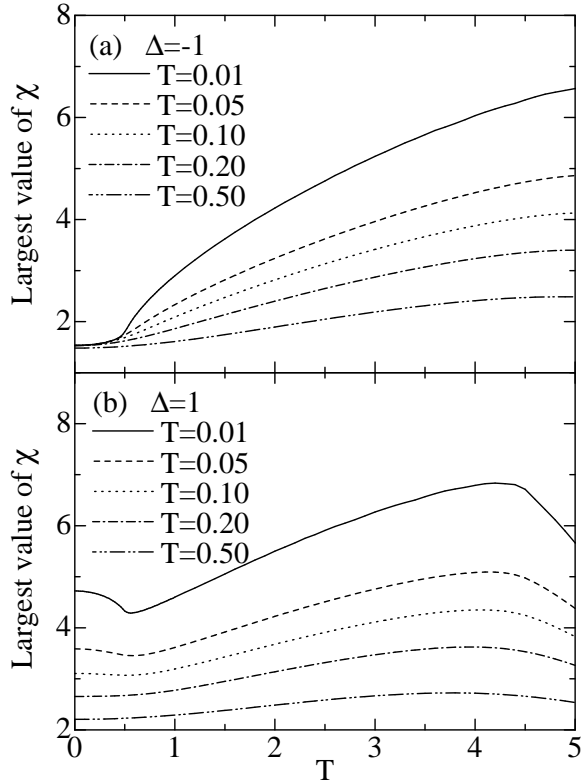}
\end{center}
\caption{Largest values of perpendicular susceptibilities $\tilde{\chi}^{\pm}_{0}$ as a function of external field for various temperature at $\gamma=1$ at (a) $\Delta=-1$ (insulator) and (b) $\Delta=1$ (semi metal)  for three-dimensional system. All wave vectors are ${\mathbf q}={\mathbf Q}$. }
\label{fig:chitemp}
\end{figure}

Let us consider the possibility of anomalous phase diagrams in the insulating and semi-metallic cases in three dimensions. 
Figure \ref{fig:chitemp} shows the temperature dependence of the perpendicular susceptibilities. 
For the insulating state, the enhancements of the perpendicular susceptibilities for the external field appear in all temperature regions at $B_{\rm ext} > |\Delta|/2$. 
On the other hand, in the semi-metallic state, the perpendicular susceptibility is sensitive to temperature. 
At low temperature ($T=0.01$), the decrease of the perpendicular susceptibility for the external field is clearly shown at $B_{\rm ext}<\Delta/2$. However, the suppressions of the perpendicular susceptibilities are weakened with an increase of the temperature. 
The amplitude is almost independent of $B_{\rm ext}$ $(<\Delta/2)$ at $T\sim 0.1$. At $T> 0.1$ the enhancement appears in overall range of the external field. 
The magnetic instability is determined by $1-U\tilde{\chi}^{\pm}_{0} \rightarrow 0$ within RPA as discussed in the previous section. 
If the magnetic order caused by the Coulomb interaction is realized at the higher temperature range which exhibits the enhancement of the perpendicular susceptibility for the external field, the anomalous $T-B$ phase diagram whose phase boundary monotonically shifts to the higher temperature region is obtained. 

\begin{figure}[tb]
\begin{center}
\includegraphics[width=8cm]{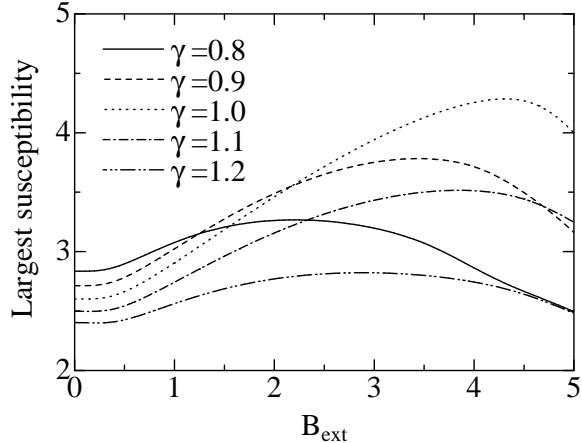}
\end{center}
\caption{Largest values of perpendicular susceptibilities $\tilde{\chi}^{\pm}_{0}$ as a function of external field for various curvature $\gamma$ at $\Delta=1$ and $T=0.1$ for three-dimensional system. All wave vectors are ${\mathbf q}={\mathbf Q}$. }
\label{fig:chigam}
\end{figure}

The effect of the curvature $\gamma$ of parabolic dispersions is shown in Fig. \ref{fig:chigam}. 
For all $\gamma$, the enhancements appear at the low field region. 
However, the amplitude of the enhancement is the most conspicuous at $\gamma=1$
 where the nesting condition survives under the presence of the external field. 
Therefore, as the curvature of one band becomes close to that of the other, it is favorable to realize the enhancement of the susceptibility for the external field.

\begin{figure}[tb]
\begin{center}
\includegraphics[width=8cm]{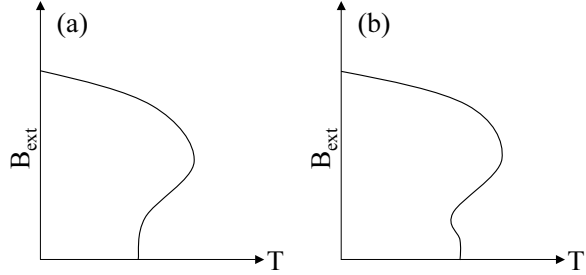}
\end{center}
\caption{Possible phase boundaries; (a) $\Delta <0$ (insulator) or $\Delta >0$ (semimetal) and higher temperature region, (b) $\Delta >0$ (semimetal) and very low temperature region ($T \ll \Delta$). }
\label{fig:apb}
\end{figure}

Finally, let us consider the structure of the phase boundary by taking account of the structures of the perpendicular susceptibilities. Figure \ref{fig:apb} shows the possible phase boundaries within RPA. 
In the insulating state ($\Delta <0$), the monotonic enhancement of the perpendicular susceptibility appears within low field region at the wide temperature region. Then, obtained phase boundary always shifts to the high temperature region with an increase of the external field (Fig. \ref{fig:apb}(a)). For the semi-metallic case ($\Delta >0$) in the higher temperature region, the structure of the perpendicular susceptibility is similar to the insulating case, so that the phase boundary shows also monotonic shift to the higher temperature region for the external field. 
Thus the phase boundary also becomes the structure of Fig. \ref{fig:apb}(a). 

On the other hand, for the semi-metallic case ($\Delta >0$) in very low temperature region, the perpendicular susceptibility becomes the concave structure within the low field region. Thus, when the magnetic order is realized in this temperature region, the phase boundary may become nontrivial structure (Fig. \ref{fig:apb}(b)).

To summarize the parabolic band model at finite temperatures, the enhancement of the perpendicular susceptibilities by the external field can appear in the insulating states with a small gap. On the other hand, for the semi-metallic states when the overlap between two bands is sufficiently small, the enhancement may appear except for very low temperature region $(T \ll \Delta)$. In very low temperature region, the perpendicular susceptibility for the external field becomes concave structure, so that the phase-boundary may become the nontrivial structure. 

In order to realize anomalous phase diagrams in the realistic semi-metallic/insulating materials, following conditions are required: 
(1) both the overlap and gap between two bands are sufficiently small; 
(2) the Fermi surfaces for electrons and holes in the semi-metallic state are very small spheres. 
(3) the change of the density of states close to the Fermi level caused by $B_{\rm ext}$ becomes conspicuous; 
(4) the curvature of one band dispersion is close to that of the other. 

\section{Summary and Discussions}
\label{sec:summary}
We have investigated the conditions for the appearance of the anomalous temperature vs. external field phase diagram in low-carrier two-band systems with a narrow gap or a small-overlap (semi-metal). 
Employing the periodic Anderson model and the parabolic dispersion model close to the Fermi level, the magnetic susceptibilities are calculated by using RPA and magnetic properties are discussed under the external field. 

For the filled-skutterudite compound CeOs$_{4}$Sb$_{12}$ whose energy dispersions may have the semi-metallic band structure, we construct the effective model by employing the periodic Anderson model with the dispersive $f$ band, which reproduces the main features of the LDA result close to the Fermi level. 
Considering magnetic properties, we show that the enhancement of the perpendicular susceptibility for the external field is related to the anomalous phase diagram observed in experimental results. 

We further introduce the two-band model with the parabolic dispersion in order to clarify the origin of the anomalous phase boundary. 
The structure of the perpendicular susceptibility for the external field is discussed in detail. 
When a small insulating gap exists in the band structure in the absence of the external field, the magnetic susceptibility perpendicular to the external field is enhanced by the small field, which results from the disappearance of the insulating gap due to the external field. On the other hand, in the semi-metallic case, the realization of anomalous phase diagrams strongly depends on the structure of density of states close to the Fermi level and the temperature region. 

We consider the effects of Coulomb interaction only by using RPA. This method is one of the mean-field approximations and neglects the effect of quantum fluctuations which may be particularly important in the vicinity of the quantum critical points.~\cite{saku05} 
Therefore, if quantum fluctuations are taken into account by employing more accurate methods, such as the fluctuation exchange (FLEX) approximation, the phase boundary may quantitatively shift to the low temperature region. 

Let us consider the effect of the orbital degeneracy. Heavy fermion compounds have the degeneracy of $f$ orbitals of rare earth and actinide ions. 
We consider the simplest model with a lower band and an upper band, where the effect of degeneracy of each band is not considered. 
If one is single band and the others have degeneracy, the perfect nesting condition is absent even at zero field. 
However, in the present study, we consider low-carrier two-band models with a small overlap and a gap between two bands, where the electron and hole surfaces become very small spheres. 
Therefore, when the Fermi spheres become obscure at finite temperature region, the nesting vector does not change even if the orbital degeneracy is considered. The structure of DOS close to Fermi level becomes most essential to the anomalous phase diagram. 
When the hole doping is also introduced, the nesting condition becomes worse. However, as far as the concentration does not conspicuously affect the number of the states near the Fermi level, the doping effect is irrelevant at finite temperature region in this study. 
Thus, we conclude that the orbital degeneracy does not change the result presented here.

\section*{Acknowledgements}
The authors would like to thank M. Kohgi, K. Iwasa, H. Harima, H. Kontani, and K. Sakurazawa for illuminating discussions. A part of the numerical computations were done at the supercomputer center at Institute for Solid State Physics, University of Tokyo. 


\end{document}